\documentclass[aps,prl,amsmath,twocolumn]{revtex4}

\usepackage{amsmath,amssymb}
\usepackage{graphicx}
\usepackage{color}
\usepackage{xspace}
\usepackage{mathtools}

\newcommand{\pder}[2]{\frac{\partial #1}{\partial #2}}

\renewcommand{\epsilon}{\varepsilon}

\preprint{Preprint}

\begin{document}

\title{Necessary Adiabatic Run Times in Quantum Optimization}

\author{Lucas~T.~Brady}
\affiliation{{Department of Physics, University of California, Santa Barbara, CA 93106-5110, USA}}
\author{Wim~van~Dam}
\affiliation{{Department of Computer Science, Department of Physics, University of California, Santa Barbara, CA 93106-5110, USA}}

\date{\today}
\begin{abstract}
Quantum annealing is guaranteed to find the ground state of optimization problems provided 
it operates in the adiabatic limit. Recent work [Phys. Rev. X 6, 031010 (2016)] has found that for some barrier tunneling problems, quantum annealing can be run much faster than is adiabatically required. 
Specifically, an $n$-qubit optimization problem was presented for which 
a non-adiabatic, or diabatic, annealing algorithm requires only constant runtime, 
while an adiabatic annealing algorithm requires a runtime polynomial in $n$. 

Here we show that this non-adiabatic speedup is a direct result of a specific symmetry in the studied problem.  In the more general case, no such non-adiabatic speedup occurs and we show why the special case achieves this speedup compared to the general case.  We also prove that the adiabatic annealing algorithm has a necessary and sufficient runtime that is quadratically better than the standard quantum adiabatic condition suggests.
We conclude with an observation about the required precision in timing 
for the diabatic algorithm. 
\end{abstract}

\maketitle


Recent work in quantum adiabatic optimization \cite{Farhi2000} 
has focused on a class of Hamming-symmetric problems that exhibits extremely strong 
non-adiabatic speedups over a slower adiabatic approach.  Numerical evidence presented by Muthukrishnan, Albash, and Lidar \cite{Muthukrishnan} shows that for several barrier tunneling problems on $n$ qubits, a well-calibrated constant time evolution of the quantum annealing Hamiltonian is sufficient. Thus, this algorithm significantly improves 
upon the slower adiabatic evolution of the Hamiltonian, which could take polynomial or even exponential time in $n$.  
Muthukrishnan et al.\ attribute this speedup to a diabatic cascade in which the ground state is quickly depopulated in favor of higher excited states and then repopulated right at the end of the diabatic evolution.

Usually the sufficient run time of quantum adiabatic optimization is estimated using the standard adiabatic condition.  This condition says that adiabaticity is ensured if the running time grows as
\begin{equation}
      \label{eq:adi_con}
      \tau\gg\max_{s\in[0,1]}\Big\|\pder{\hat{H}(s)}{s}\Big\|/g(s)^2,
\end{equation}
for the spectral gap $g(s)$.  More accurate versions of this condition have been proven \cite{Jansen}, but all of them depend linearly on the matrix norm of $\hat{H}(s)$ or its derivatives with respect to $s$ divided by a low degree polynomial function of the gap $g(s)$.  

The condition in Eq.~\ref{eq:adi_con} is merely a sufficient condition, and it is possible to have adiabatic evolutions with 
shorter running times than Eq.~\ref{eq:adi_con} describes. 
Furthermore it is also possible to have a non-adiabatic evolution that succeeds 
in solving the optimization problem at hand. It is such a 
non-adiabatic speedup that is described by Muthukrishnan et al.~\cite{Muthukrishnan}.

A non-adiabatic speedup is obviously significant for near-term quantum computers where quantum annealing is a potential application.  
Kong and Crosson \cite{Kong} have studied these diabatic transitions, and more recently the current authors presented complementary findings \cite{AQC}.  These recent results indicates that this non-adiabatic speedup can provide an alternate and efficient way of solving an important class of Hamming-symmetric barrier tunneling problems that are being used as toy models \cite{Muthukrishnan,Kong,Farhi2002,Reichardt,Crosson,Harrow,Jiang,Brady2} to study the more general properties of quantum annealing in the presence of a barrier.

Here we present results that indicate that even slightly more generalized versions of symmetric barrier tunneling problems do not exhibit this fast non-adiabatic speedup.  The base Hamiltonian used to study this class of problems exists in a Hilbert space of $n$ qubits and is given by
\begin{equation}
      \hat{H}(s) = -(1-s)\sum_{i=1}^n \sigma_x^{(i)} + s\left[\sum_{i=1}^n \sigma_z^{(i)}+ b\left(\sum_{i=1}^n \sigma_z^{(i)}\right)\right],
\end{equation}
where $b(h)$ is some localized barrier or perturbation and $s = t/\tau$ is a normalized time variable representing the linear progression of time, $t$, from $t=0$ to the algorithm stopping time $\tau$.  Current numerical evidence \cite{Muthukrishnan} suggests that the non-adiabatic speedup exists for many classes, shapes, and sizes of localized barriers $b(h)$.  This article generalizes the problem slightly (ignoring $b(h)$ for the moment):
\begin{equation}
      \hat{H}(s) = -(1-s)\sum_{i=1}^n \sigma_x^{(i)} + s\mu\sum_{i=1}^n \sigma_z^{(i)},
\end{equation}
by introducing a positive slope parameter $\mu$
and we find that for the generic case $\mu\neq 1$, the non-adiabatic speedup no longer exists.
We call $\mu$ a slope as it relates linearly the energy of the system with the Hamming weight 
$\sum_i \sigma_z^{(i)}$ of the $n$ qubits.  

Since this Hamiltonian describes a simple toy model, it is unlikely that a physical system will exhibit the exact $\mu=1$ behavior, leading us to the conclusion that for realizable problems, this diabatic speedup will not exist.  In this article, we will focus on the $b(h)=0$ case since it decouples all the qubits, allowing us to extract information about the system by studying the evolution of a single qubit Hamiltonian.  Since $\mu\neq1$ disrupts the non-adiabatic speedup even in this $b(h)=0$ case, we fully expect similar disruption to occur for more complicated barriers and perturbations.


We first need to define our criteria for an optimal runtime.  If an algorithm on $n$ qubits runs for time $\tau$ and has a probability of success of $p_n(\tau)$ at the end of that time, its expected running time is $\tau/p_n(\tau)$, and the optimal running time is the $\tau_n$ that minimizes $\tau/p_n(\tau)$ for $n$ qubits.  In our case, we have $n$ independent qubits, each of which has a probability of success of $p_1$, hence $p_n = p_1^n$, which is where the $n$ dependence comes into the minimization.

\begin{figure}
      \begin{center}
            \includegraphics[width=0.48\textwidth]{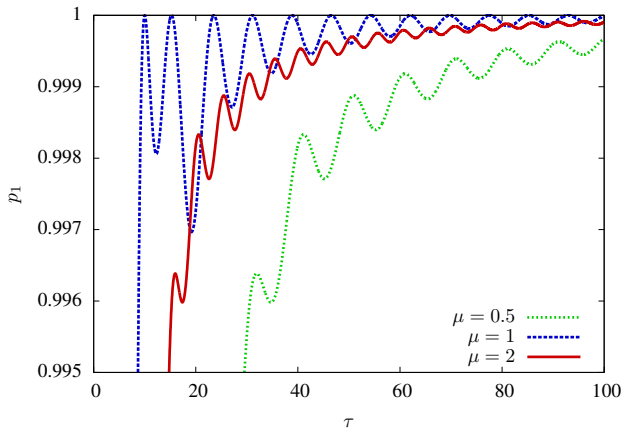}
      \end{center}
      \caption{
            The single qubit success probability, $p_1$, as a function of the total runtime for several $\mu$ values.  The blue, dashed, $\mu=1$ line corresponds to the model that has been studied in previous articles.  Notice that the $\mu=1$ curve has several special properties, including that it goes to $p_1=1$ at finite $\tau$, resulting in the non-adiabatic speedup noted in other papers.  The $\mu\neq1$ curves do not exhibit this $p_1=1$ behavior.
      }
      \label{fig:pf_vs_tau}
\end{figure}

In the $\mu=1$ case, $p_1$ goes to $1$ for finite $\tau$, as seen in Fig.~\ref{fig:pf_vs_tau}, meaning that $p_n=1$ at this value, leading to the non-adiabatic speedup noted in other studies.  Fig.~\ref{fig:pf_vs_tau} also shows $\mu=0.5$ and $\mu=2$ curves.  Note that for these curves the success probability does not achieve $p_1=1$  at finite $\tau$.  Similar plots can be obtained for other $\mu\neq 1$ and, as we note below, this failure to reach $p_1=1$ for finite $\tau$, ultimately leads to the breaking of the non-adiabatic speedup.  Therefore, this speedup is restricted to the special case of $\mu=1$.

\begin{figure}
      \begin{center}
            \includegraphics[width=0.48\textwidth]{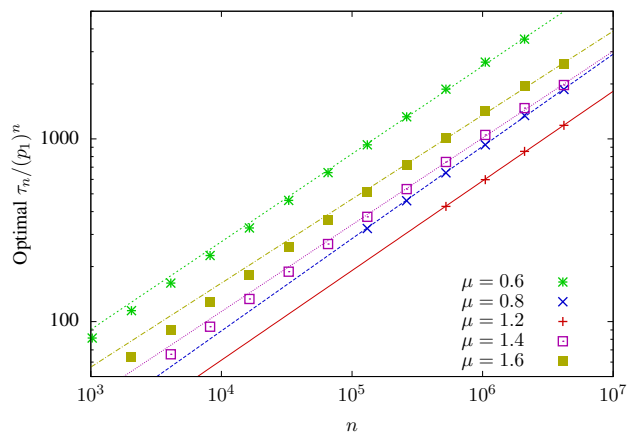}
      \end{center}
      \caption{
            Optimal expected running time of quantum annealing, $\tau_n/p_n(\tau_n)$, as a function of $n$ for different $\mu$ values.  Unlike the $\mu=1$ case, $\tau_n$ increases with $n$ for these $\mu$ values.  lines through the data are power law fits of the form $\tau_n = A n^r$, and the fitted $r$ values in the order $\mu=(0.6,0.8,1.2,1.4,1.6)$ are $(0.48,0.51,0.49,0.48,0.46)$, all close to $1/2$.  A scaling power of $1/2$ is consistent with the adiabatic scaling of the $\mu=1$ case as found in \cite{Muthukrishnan} and our results below while being quadratically faster than the sufficient adiabatic condition.
      }
      \label{fig:tau_vs_n}
\end{figure}

To demonstrate the lack of a non-adiabatic speedup in the $\mu\neq1$ cases, consider Fig.~\ref{fig:tau_vs_n},  which shows the optimal expected runtime, $\tau_n/p_n(\tau_n)$, as a function of $n$.  All of the $\mu$ curves shown are increasing, meaning that the running time increases with $n$, and there is no non-adiabatic algorithm that runs in constant time.  The fitted curves are to power laws of the form $\tau_n = A n^p$, and all of the fitted $p$ values are close to $1/2$, indicating a running time of $\mathcal{O}(\sqrt{n})$.


\begin{figure}
      \begin{center}
            \includegraphics[width=0.48\textwidth]{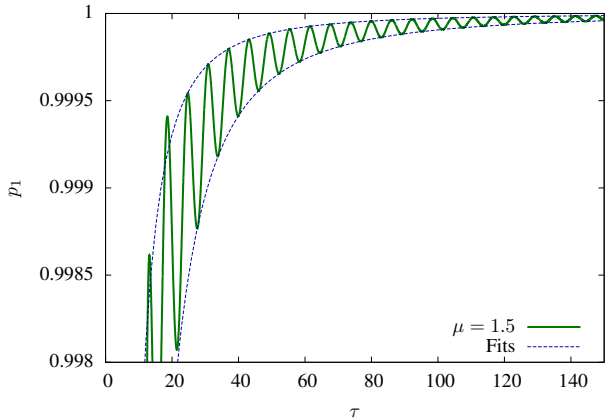}
      \end{center}
      \caption{
            A single qubit success probability curve as a function of total runtime $\tau$ for $\mu=1.5$ with upper and lower envelopes shown.  These envelopes were obtained by first extracting all the local minima (maxima) and doing a power law fit of the form $p_1 = 1 - c \tau^{-q}$.  The first two minima and maxima were excluded from this fit and others since they tend to be more abnormal.  In this case, the upper envelope has a fitted $q=1.998$ and the lower envelope has a fitted $q=1.996$, both of which are extremely close to the $2$ we use in the text.
      }
      \label{fig:envelopes}
\end{figure}

We can extract the $\sqrt{n}$ running time behavior from the curves in Fig.~\ref{fig:pf_vs_tau} as well
because the qubits in our problem are completely decoupled.
For sufficiently large running times $\tau$, the curves of the single qubit success probability $p_1$ as a function of $\tau$ shown in Fig.~\ref{fig:pf_vs_tau} are bounded above and below by envelopes of the form
\begin{equation}
      \label{eq:ineq}
      1- \frac{c_{\ell}(\mu)}{\tau^q}<p_1 < 1-\frac{c_u(\mu)}{\tau^q}, 
\end{equation}
with constants $c_{\ell}(\mu)$ and $c_u(\mu)$.  
This relationship is extracted by performing numerical fits to the minima and maxima in curves like those seen in Fig.~\ref{fig:envelopes}, and for all our fits to different $\mu$ data, $q$ is close to $2$.  Note that $c_u(1)=0$, which, as we will see, is one of the main reasons why the $\mu=1$ diabatic speedup can occur.

Muthukrishnan et al. \cite{Muthukrishnan} showed that the lower envelope with $c_{\ell}(\mu)$ guarantees that the worst case running-time for the $\mu=1$ case scales as $\mathcal{O}(\sqrt{n})$.  We will employ their method to show that a relationship such as Eq.~\ref{eq:ineq} provides both the necessary and sufficient condition for the running time.  Muthukrishnan et al.\ also apply methods created by Boixo and Somma \cite{Boixo} to show that at least $\Omega(n^{1/2})$ is necessary for adiabatic evolution.

If for $n$ qubits a total success probability of $p$ is desired from the algorithm, then Eq.~\ref{eq:ineq} tells us that
\begin{equation}
      \label{eq:envelope_1}
      \left(1- \frac{c_{\ell}(\mu)}{\tau^q}\right)^n \leq p \leq \left(1-\frac{c_u(\mu)}{\tau^q}\right)^n.
\end{equation}

We can manipulate this inequality, performing an expansion for small $c_*(\mu)/\tau^q$ since $\tau$ will be large.  The result of these manipulations gives us a relationship between the running time and $n$:
\begin{equation}
      \label{eq:envelope_2}
      \left(\frac{c_u(\mu)}{\ln 1/p}\,n\right)^{1/q}      
      \leq  \tau \leq 
      \left(\frac{c_{\ell}(\mu)}{\ln 1/p}\,n\right)^{1/q}.
\end{equation}
Therefore, since $q=2$ in our cases, having a running time that scales as $\sqrt{n}$ is both a necessary and sufficient condition to reaching a desired probability.  Note that when $\mu=1$, $c_u(1)=0$, so one side is no longer bounded, leading to the possibility of a non-adiabatic speedup.

In the Hamming weight problem, the gap is constant with $n$, and all matrix norms of the Hamiltonian and its derivatives will depend linearly on $n$.  Therefore, the adiabatic condition, Eq.~\ref{eq:adi_con}, would predict $\mathcal{O}(n)$ scaling; whereas, our results indicate that a faster $\mathcal{O}(\sqrt{n})$ running time is sufficient.  This result was shown in \cite{Muthukrishnan} for $\mu=1$, and our results indicate that this quadratic speedup holds for general slopes $\mu$. 

While the standard adiabatic condition overestimates the running time, there are other derivations that apply to our problem more specifically 
and that provide a stricter bound that matches our results.  
Jansen, Ruskai, and Seiler \cite{Jansen} showed that for 
fixed Hamiltonians $\hat{H}_0$ and $\hat{H}_1$ with time evolution 
 $\hat{H}(t) = (1-t/\tau)\hat{H}_0+t/\tau\hat{H}_1$, the 
 success probability $p$ of remaining in the ground state throughout 
 $0\leq t \leq \tau$ is  bounded by
\begin{equation}
      \label{eq:Jansen}
      p = 1-\mathcal{O}(\tau^{-2}).
\end{equation}
If we take this to be the probability of success for a single qubit case, our results in Eqs.~\ref{eq:envelope_1} and \ref{eq:envelope_2} imply 
that $\tau\in\mathcal{O}(\sqrt{n})$ is sufficient for an adiabatic 
evolution.  This shows that the result from Jansen et al.\ provides a stricter sufficient condition than the standard adiabatic condition for our 
optimization problem with decoupled qubits.

\begin{figure}
      \begin{center}
            \includegraphics[width=0.48\textwidth]{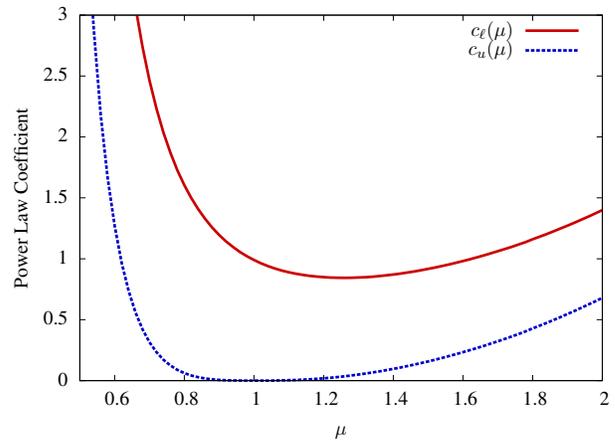}
      \end{center}
      \caption{
            The curves like the one in Fig.~\ref{fig:envelopes} are bounded above and below by curves of the form $1-c/\tau^2$.  We show the values of $c$ for the upper, $c_u$, and lower, $c_{\ell}$, bounding functions as obtained from numerical fits.  These coefficients are a function of $\mu$, and all of the fits used to obtain this data were good quality.  In the main text, we show that these bounding curves directly lead to a $\mathcal{O}(\sqrt{n})$ running time for the algorithm in all cases except the $\mu=1$ case where $c_u(1)=0$.
      }
      \label{fig:c_vs_mu}
\end{figure}

In Fig.~\ref{fig:c_vs_mu} we plot the coefficients 
$c_u(\mu)$ and $c_{\ell}(\mu)$ obtained from numerical fits.  The fits used to obtain these values are akin to those shown in Fig.~\ref{fig:envelopes}, making us confident in the $1/\tau^2$ scaling of the error.  Notice that as we approach the special case $\mu=1$ we see that $c_u(\mu)\to0$ and observe that around 
$\mu=1$ the coefficient $c_u(\mu)$ stays close to zero. 
Hence for $\mu$ approximately (but not exactly) $1$, the non-adiabatic speedup will persists for a large range of $n$ until the adiabatic running time 
of $\mathcal{O}(\sqrt{n})$ is required again at very large $n$.

\begin{figure}
      \begin{center}
            \includegraphics[width=0.48\textwidth]{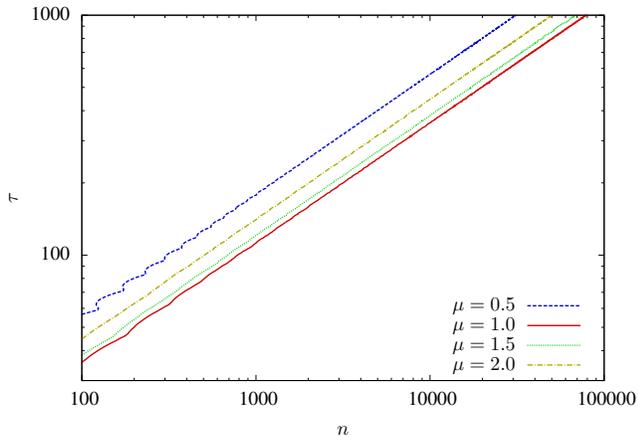}
      \end{center}
      \caption{
            These plots shows the runtime, $\tau$, needed to ensure that state of the system is at least $75\%$ in the ground state over the entire $s$ evolution.  This growth of $\tau$ with $n$ comes closest to a true adiabatic evolution, and we can see that the $\tau\in\mathcal{O}(\sqrt{n})$ behavior holds even in this case.  Power law fits to these data sets show that the exponent for these curves, in the order $\mu = (0.5,1.0,1.5,2.0)$, are $(0.497,0.502,0.501,0.500)$.  Therefore, the quadratic speedup we see over the sufficient adiabatic condition is a property of adiabatic evolution in this system, not the specific $\tau/p_n$ criteria we used.
      }
      \label{fig:true_adiabatic}
\end{figure}

All of our work so far has shown that the optimal running time of this algorithm is $\mathcal{O}(\sqrt{n})$, but this does not imply that the optimal running time results from adiabatic evolution.  If we look at the occupancy of the energy states for these optimal runs, we in fact see the ground state being depopulated during the $s$ evolution.  Therefore, a remaining question to ask is whether this behavior also holds if we require the system to stay within a certain range of its ground state for the entire $s\in[0,1]$ evolution.

In Fig.~\ref{fig:true_adiabatic}, we show the time, $\tau$, needed to ensure that the system has at least a $75\%$ chance of being measured in its ground state for the entire $s\in[0,1]$ evolution.  All of these curves exhibit power law relationships, $\tau = B n^r$, with fitted $r=(0.497,0.502,0.501,0.500)$ for $\mu = (0.5,1.0,1.5,2.0)$ respectively.  A similar plot can be obtained if a stricter cutoff than $75\%$ is used.

Fig.~\ref{fig:true_adiabatic} shows that the runtime relationships we observe are in fact indicative of how adiabatic evolution behaves as well.  Therefore, we are led to the conclusion that for general $\mu\neq1$, the runtime $\tau\in\Theta(\sqrt{n})$ is both necessary and sufficient to ensure finding the true ground state.  The $\mu=1$ case remains a special case that goes against this rule, allowing for an extreme speedup to a constant running time.


Our last goal will be to understand the width of the success probability spike of $p_1$ in the unperturbed, $\mu=1$ case when it reaches the optimal $p_1=1$.
We will show that this narrowness implies that to be successful for large $n$, one has to be very precise in using the right running time $\tau$. 

We know that there is a critical runtime $\tau_c$ such that $p_1=1$ for a single qubit.  For run times close to this $\tau_c$, the probability of success can be modeled by
\begin{align}
      p_1 = 1-\delta= 1-k(\tau-\tau_c)^{2},\qquad \delta\ll 1,
\end{align}
where $|\tau-\tau_c|$ is the required stopping precision of the algorithm.

Scaling the system to $n$ qubits, the probability of success is $p_n=p_1^n$ since the qubits are uncoupled in the unperturbed case:
\begin{align}
      p_n = \left(1-k(\tau-\tau_c)^{2}\right)^n \approx 1-nk(\tau-\tau_c)^{2}.
\end{align}
If we want the probability of failure to be less than $\epsilon$, 
we must have that 
\begin{equation}
      1-\epsilon <1-kn(\tau-\tau_c)^2 \Rightarrow |\tau-\tau_c| < (\epsilon/kn)^{1/2}.
\end{equation}
Thus, maintaining the same success probability as $n$ increases requires the acceptable imprecision 
$|\tau-\tau_c|$ to shrink according to $n^{-1/2}$.  

For perturbed problems with a barrier we have run simulations using square barriers like those considered in \cite{Brady2}. For $\mu=1$, we found the same $n^{-1/2}$ narrowing of the spiked success probability $p_n$
around the critical $\tau_c$ running time.



Our conclusion is therefore that, while the $\mu=1$ case does exhibit a surprising non-adiabatic speedup that could potentially be exploited, this diabatic speedup is not a general feature of this class of quantum annealing problems.  Running these algorithms adiabatically remains the best and only option to achieve success in general.

\subsubsection*{Acknowledgements}
This material is based upon work supported by the National Science Foundation under Grants No.\ 1314969 and No.\ 1620843.


\end{document}